\documentclass[11pt]{article}
\usepackage{amsmath,amssymb,epsf}

\def\nn{\nonumber}

\let\bm=\bibitem

\newcommand{\be}{\begin{equation}}
\newcommand{\ee}{\end{equation}}
\def\ba{\begin{array}}
\def\ea{\end{array}}

\def\fft#1#2{\frac{#1}{#2}}

\def\sst#1{{\scriptscriptstyle #1}}

\def\td{\tilde}

\def\dalemb#1#2{{\vbox{\hrule height .#2pt
        \hbox{\vrule width.#2pt height#1pt \kern#1pt
                \vrule width.#2pt}
        \hrule height.#2pt}}}

\newcommand{\hoch}[1]{$\, ^{#1}$}
\newcommand{\bea}{\begin{eqnarray}}
\newcommand{\eea}{\end{eqnarray}}

\def\0{{\sst{(0)}}}
\def\1{{\sst{(1)}}}
\def\2{{\sst{(2)}}}
\def\3{{\sst{(3)}}}
\def\4{{\sst{(4)}}}
\def\5{{\sst{(5)}}}
\def\6{{\sst{(6)}}}
\def\7{{\sst{(7)}}}
\def\8{{\sst{(8)}}}

\begin{document}

\begin{flushright}
MIFP-07-15 \\
{\bf arXiv:0705.4471}\\
May\  2006
\end{flushright}

\begin{center}

{\Large {\bf Kerr-Schild Structure and Harmonic 2-forms on \\
(A)dS-Kerr-NUT Metrics}}

\vspace{20pt}

W. Chen and H. L\"u

\vspace{20pt}

{\hoch{\dagger}\it George P. \&  Cynthia W. Mitchell Institute
for Fundamental Physics,\\
Texas A\&M University, College Station, TX 77843-4242, USA}

\vspace{40pt}

\underline{ABSTRACT}
\end{center}

We demonstrate that the general (A)dS-Kerr-NUT solutions in $D$
dimensions with $([D/2], [(D+1)/2])$ signature admit $[D/2]$
linearly-independent, mutually-orthogonal and affinely-parameterised
null geodesic congruences.  This enables us to write the metrics in a
multi-Kerr-Schild form, where the mass and all of the NUT parameters
enter the metrics linearly.  In the case of $D=2n$, we also obtain
$n$ harmonic 2-forms, which can be viewed as charged
(A)dS-Kerr-NUT solution at the linear level of small-charge expansion,
for the higher-dimensional Einstein-Maxwell theories.  In the BPS
limit, these 2-forms reduce to $n-1$ linearly-independent ones,
whilst the resulting Calabi-Yau metric acquires a K\"ahler 2-form,
leaving the total number the same.



\newpage

\section{Introduction}

     One intriguing feature of General Relativity is that, despite its
high degree of non-linearity, many exact solutions can be cast into a
Kerr-Schild form \cite{ksform} where non-trivial parameters such as
mass, charge, or cosmological constant enter the metrics as a linear
perturbation of flat spacetime.  A simple example is the (A)dS metric,
which can be written as
\be
ds^2=-dt^2 + dr^2 + r^2 d\Omega_n^2 + \Lambda\, r^2 (dt-dr)^2\,,
\ee
where the first three terms describe the $(n+2)$-dimensional
Minkowski spacetime and the cosmological constant enters the last term
linearly.  More complicated examples include the Plebanski metric
\cite{pleb}; in (2,2) signature, the Plebanski metric can have a
double Kerr-Schild form where both the mass and the NUT charge enter
the metric linearly \cite{cglp}.

       The most general higher-dimensional (A)dS-Kerr-NUT solutions,
which can be viewed as higher-dimensional generalisations of the
Plebanski metric, were recently obtained in \cite{clpnut}.  The
solutions are parameterised by the mass, multiple NUT charges and
arbitrary orthogonal rotations. The metrics have $U(1)^n$ isometries,
where $n=[(D+1)/2]$.  They are demonstrated \cite{curvature} to be of
type D in the higher-dimensional generalisation \cite{classification}
of the Petrov classification.

           Many further interesting properties of the metrics were
obtained, such as the separability of the Hamiltonian-Jacobi and
Klein-Gordon equations \cite{separability}, and the existence of
Killing-Yano tensors \cite{killingyano}.  The metrics also admit BPS
limits where the Killing spinors can emerge \cite{clpnut}.  In the odd
$2n+1$ dimensions, this leads to a large class of Einstein-Sasaki
metrics with $U(1)^n$ isometry, generalising the previously known
$Y^{p,q}$ \cite{ypq} and $L^{pqr}$ \cite{lpqr} spaces.  In the even
$2n$ dimensions, this leads to the non-compact Calabi-Yau metrics that
can provide a resolution of the cone over the Einstein-Sasaki
metrics constructed in the odd dimensions \cite{res1,res2}.

       In this letter, we demonstrate in section 2 that the
$D$-dimensional (A)dS-Kerr-NUT solution admits $[D/2]$
linearly-independent, mutually-orthogonal and affinely parameterised
null geodesic congruences upon Wick-rotation of the metric to $([D/2],
[(D+1)/2])$ signature.  This enables us to cast the metric into the
multi-Kerr-Schild form, where the mass and all of the NUT parameters
enter the metric linearly.  In section 3, we obtain $n$ harmonic
2-forms on the (A)dS-Kerr-NUT metrics in $D=2n$ dimensions.  In the
BPS limit, these $n$ harmonic 2-forms becomes linearly dependent, and
the number of linearly-independent ones becomes $n-1$.  However, a
K\"ahler 2-form emerges under the BPS limit, and hence the
total number of harmonic 2-forms remains $n$.  We conclude the letter
in section 4.

\section{Multi-Kerr-Schild structure}

Let us first consider the case of $D=2n+1$ dimensions, for which the
metric was given in \cite{clpnut}.  In order to put the metric in a
Kerr-Schild form, it is necessary to Wick rotate to $(n,n+1)$
signature.  This can be easily achieved by Wick rotating all the
spatial $U(1)$ coordinates.  The corresponding metric is then given by
\be ds^2= \sum_{\mu=1}^n \Big\{
         \fft{dx_\mu^2}{Q_\mu} - Q_\mu\, \Big( \sum_{k=0}^{n-1}
          A_\mu^{(k)}\, d\psi_k\Big)^2\Big\} +
           \fft{c}{(\prod_{\nu=1}^n x_\nu^2)}\,
           \Big( \sum_{k=0}^n A^{(k)} \, d\psi_k\Big)^2
\,,\label{2n1pleb} \ee
where
\bea Q_\mu &=& \fft{X_\mu}{U_\mu}\,,\qquad U_\mu =
{{\prod}'}_{\nu=1}^n
     (x_\nu^2 - x_\mu^2)\,, \qquad X_\mu = \sum_{k=1}^n c_k\, x_\mu^{2k}
         + \fft{c}{x_\mu^2}  - 2 b_\mu\,,\nn\\
A_\mu^{(k)} &=& \sum'_{\nu_1 <\nu_2 <\cdots < \nu_k}
            x_{\nu_1}^2 x_{\nu_2}^2\cdots
        x_{\nu_k}^2\,,\qquad A^{(k)} = \sum_{\nu_1<\nu_2\cdots <\nu_k}
               x_{\nu_1}^2 x_{\nu_2}^2 \cdots x_{\nu_k}^2\,,\label{QY}
\eea
The prime on the summation and product symbols in the definition
of $A_\mu^{(k)}$ and $U_\mu$
indicates that the index value $\mu$ is omitted in the summations of
the $\nu$ indices over the range $[1,n]$. Note that $\psi_0$ was
denoted as $t$ in \cite{clpnut}, playing the r\^ole of the time like
coordinate in the $(1, 2n)$ spacetime signature.  In this way of
writing the metric, all of the integration constants of the solution
enter only in the functions $X_\mu$.  The constant $c_n=(-1)^{n}
\Lambda$ is fixed by the value of the cosmological constant, with
$R_{\mu\nu}=2n \Lambda\, g_{\mu\nu}$.  The other $2n$ constants $c_k$, $c$
and $b_\mu$ are arbitrary. These are related to the $n$ rotation
parameters, the mass and the $(n-1)$ NUT parameters, with one
parameter being trivial and removable through a scaling symmetry
\cite{clpnut}.  Note that in $(n,n+1)$ signature, the NUT charges are
really masses with respect to different time-like Killing vectors.
However, we shall continue to refer them as NUT charges.

     We now re-arrange the metric (\ref{2n1pleb}) into the form
\bea ds^2 &=& - \sum_{\mu=1}^n \fft{X_\mu}{U_\mu}
          \Big[  \sum_{k=0}^{n-1}
          A_\mu^{(k)}\, d\psi_k + \fft{U_\mu}{X_\mu}dx_{\mu} \Big]\,
          \Big[  \sum_{k=0}^{n-1}
          A_\mu^{(k)}\, d\psi_k - \fft{U_\mu}{X_\mu}dx_{\mu} \Big]
\nn\\
&&   + \fft{c}{(\prod_{\nu=1}^n x_\nu^2)}\,
           \Big( \sum_{k=0}^n A^{(k)} \, d\psi_k\Big)^2\,. \eea
If we perform the following coordinate transformation,
\be d\hat{\psi}_k = d\psi_k + \sum^n_{\mu=1}
                 \fft{(- x_\mu^2)^{n-k-1}}{X_{\mu}}dx_{\mu}\,,
                 \qquad k \, = \, 0 \,,\:\cdots\:,n\,,
\ee
the metric can then be cast into the n-Kerr-Schild form,
namely
\be ds^2 = d\bar s^2 + \sum_{\mu=1}^n\fft{2b_{\mu}}{U_{\mu}}
           \Big[\sum_{k=0}^{n-1} A_\mu^{(k)}\, d\hat \psi_k \Big]^2\,,
\label{oddks1}
\ee
where
\bea  d\bar s^2 &=&  - \sum_{\mu=1}^n \Big\{\fft{\bar X_\mu}{U_\mu}
          \Big[  \sum_{k=0}^{n-1}
          A_\mu^{(k)}\, d\hat \psi_k \Big]^2
          - 2 \Big[  \sum_{k=0}^{n-1}
          A_\mu^{(k)}\, d\hat \psi_k  \Big]dx_{\mu}\Big\}\nn\\
&& + \fft{c}{(\prod_{\nu=1}^n x_\nu^2)}\,
           \Big( \sum_{k=0}^n A^{(k)} \, d\hat \psi_k\Big)^2\,,\nn\\
   \bar X_\mu &=& \sum_{k=1}^n c_k\, x_\mu^{2k}
         + \fft{c}{x_\mu^2}\,.\label{adsmetric1}
\eea
It is straightforward to verify that the metric $d\bar s^2$ is that of
pure (A)dS spacetime.  The mass and NUT parameters $b_\mu$ appear
linearly in the metric $ds^2$.   It should be emphasised that
although the constants $c$ and $c_k$ with $k<n$ are trivial
in the metric $d\bar s^2$, they provide non-trivial angular
momentum parameters in the metric $ds^2$. It is interesting to note
that all of the constants $c_k$, including $c_n$ that is related to the
cosmological constant, appear linearly in the metric, and can all be
extracted from $d\bar s^2$ and grouped in the second term of
(\ref{oddks1}).  This implies that all the parameters, the mass,
NUTs and angular momenta and cosmological constant can enter
the metric linearly as a perturbation of flat spacetime.  In this letter,
we shall consider in detail only the Kerr-Schild form where the
mass and NUT parameters enter the metric linearly as a perturbation of
pure (A)dS spacetime.

         The (A)dS metric (\ref{adsmetric1}) can be diagonalised, in a
way that the second term of (\ref{oddks1}) remains simple.  To
do so, let us first rewrite the $\bar X_\mu$ as follows
\be
\bar X_\mu=\fft{(1 + \Lambda\,x_\mu^2)}{x_\mu^2}
\prod_{k-1}^n(a_k^2 - x_\mu^2)
\,.
\ee
Then we complete the square in $d\bar s^2$:
\be  d\bar s^2 =   \sum_{\mu=1}^n \Big\{
            \fft{U_\mu}{\bar X_\mu}dx_{\mu}^2
          - \fft{\bar X_\mu}{U_\mu}
            \Big[  \sum_{k=0}^{n-1}
           A_\mu^{(k)}\, d\hat \psi_k
           - \fft{U_\mu}{\bar X_\mu} dx_{\mu}\Big]^2\Big\}
            + \fft{c}{(\prod_{\nu=1}^n x_\nu^2)}\,
           \Big( \sum_{k=0}^n A^{(k)} \, d\hat \psi_k\Big)^2\,,
\ee
and make the coordinate transformation,
\be d\td\psi_k = - d\hat \psi_k + \sum^n_{\mu=1}
                 \fft{(- x_\mu^2)^{n-k-1}}{\bar X_{\mu}}dx_{\mu}\,,
                 \qquad k \, = \, 0 \,,\:\cdots\:,n\,.
\ee
The metric can be put into a new form,
\be ds^2 = d\bar s^2 + \sum_{\mu=1}^n\fft{2b_{\mu}}{U_{\mu}}
           \Big[\sum_{k=0}^{n-1} A_\mu^{(k)}\, d\td \psi_k
           - \fft{U_\mu}{\bar X_\mu} dx_{\mu}\Big]^2\,,
\ee
where
\be  d\bar s^2 =   \sum_{\mu=1}^n \Big\{
            \fft{U_\mu}{\bar X_\mu}dx_{\mu}^2
          - \fft{\bar X_\mu}{U_\mu}
            \Big[  \sum_{k=0}^{n-1}
           A_\mu^{(k)}\, d\td \psi_k \Big]^2\Big\}
            + \fft{c}{(\prod_{\nu=1}^n x_\nu^2)}\,
           \Big( \sum_{k=0}^n A^{(k)} \, d\td \psi_k\Big)^2\,.
\ee
Performing a recombination of the $U(1)$ coordinates, namely
\be     \tau =  \sum_{k=0}^n B^{(k)} \, d\td \psi_k\,, \qquad
  \fft{\varphi_i}{a_i} = \sum^n_{k=1}B^{(k-1)}_i d\td\psi_k
                 -\Lambda \sum_{k=0}^{n-1} B_i^{(k)}\, d\td \psi_k\,,
                \qquad i \, = \, 1 \,,\:\cdots\:,n\,,
           \ee
where
\be B_i^{(k)} = \sum'_{j_1 <j_2 <\cdots < j_k}
            a_{j_1}^2 a_{j_2}^2\cdots
        a_{j_k}^2\,,\qquad
        B^{(k)} =
         \sum_{j_1<j_2\cdots <j_k}
               a_{j_1}^2 a_{j_2}^2 \cdots a_{j_k}^2\,,
\ee
the odd dimensional (A)dS-Kerr-NUT metrics can be expressed as
\be ds^2 = d\bar{s}^2 + \sum_{\mu=1}^{n}
          \fft{2b_{\mu}}{U_{\mu}}(k_{(\mu)\alpha}dx^{\alpha})^2\,,
\ee
\bea d\bar{s}^2 &=&  \fft{W}{\prod^n_{i=1} \Xi_i} d\tau^2
            + \sum_{\mu=1}^{n} \fft{U_\mu}{\bar X_\mu}\, dx_\mu^2
            - \sum_{i=1}^n \fft{\gamma_i}
            {\Xi_i {{\prod}'}_{k=1}^{\, n} (a_i^2-a_k^2)}
            d\varphi_i^2\,, \\
    k_{(\mu)\alpha}dx^{\alpha} &=& \fft{W}{1+\Lambda\, x_\mu^2}
                \fft{d\tau}{\prod^n_{i=1} \Xi_i}
                - \fft{U_{\mu} \, dx_{\mu}} {\bar X_{\mu}}
                - \sum_{i=1}^n \fft{a_i \, \gamma_i d\varphi_i}
                {(a_i^2- x_\mu^2)\Xi_i {{\prod}'}_{k=1}^{\, n}
(a_i^2-a_k^2)}\,,
\eea
where
\be
\Xi_i=1+\Lambda\, a_i^2\,,\qquad
\gamma_i=\prod_{\nu=1}^n(a_i^2-x_\nu^2)\,,\qquad
W=\prod_{\nu=1}^n (1 + \Lambda\, x_\nu^2)\,.\label{extrastr}
\ee
If we set all but one of the $b_\mu$ to zero, the result reduces to
the Kerr-Schild form for rotating (A)dS black holes obtained
previously in \cite{glpp}.

     We now turn our attention to the the case of $D=2n$ dimensions.
The corresponding (A)dS-Kerr-NUT metrics were obtained in
\cite{clpnut}.  After performing Wick rotations, the metric with
$(n,n)$ signature is given by
\be ds^2= \sum_{\mu=1}^n \Big\{
         \fft{dx_\mu^2}{Q_\mu} - Q_\mu\, \Big( \sum_{k=0}^{n-1}
          A_\mu^{(k)}\, d\psi_k\Big)^2\Big\}\,,\label{2npleb}
\ee
where we $Q_\mu$, $U_\mu$ and $A_\mu^{(k)}$ have the same form as
those in the even dimensions, given in (\ref{QY}).  The functions
$X_\mu$ are given by
\be
X_\mu = \sum_{k=0}^n c_k\, x_\mu^{2k} +
      2 b_\mu\, x_\mu\,.\label{QYeven}
\ee
The constants
$c_k$ and $b_\mu$ are arbitrary, except for $c_n=(-1)^{n} \Lambda$, which
is fixed by the value of the cosmological constant, $R_{\mu\nu}
=(2n-1)\Lambda\, g_{\mu\nu}$.
The metric can be re-arranged into the form
\be ds^2 = - \sum_{\mu=1}^n \fft{X_\mu}{U_\mu}
          \Big[  \sum_{k=0}^{n-1}
          A_\mu^{(k)}\, d\psi_k + \fft{U_\mu}{X_\mu}dx_{\mu} \Big]\,
          \Big[  \sum_{k=0}^{n-1}
          A_\mu^{(k)}\, d\psi_k - \fft{U_\mu}{X_\mu}dx_{\mu} \Big]
\,.
 \ee
After performing the coordinate transformation
\be d\hat{\psi}_k = d\psi_k + \sum^n_{\mu=1}
                 \fft{(- x_\mu^2)^{n-k-1}}{X_{\mu}}dx_{\mu}\,,
                 \qquad k \, = \, 0 \,,\:\cdots\:,n-1\,,
\ee
the metric can be cast into the $n$-Kerr-Schild form,
\be ds^2 = d\bar s^2 -
\sum_{\mu=1}^n\fft{2b_{\mu}x_{\mu}}{U_{\mu}}
           \Big[\sum_{k=0}^{n-1} A_\mu^{(k)}\, d\hat \psi_k \Big]^2\,
\label{evenks1}
\ee
where
\bea  d\bar s^2 &=&  - \sum_{\mu=1}^n \Big\{\fft{\bar X_\mu}{U_\mu}
          \Big[  \sum_{k=0}^{n-1}
          A_\mu^{(k)}\, d\hat \psi_k \Big]^2
          - 2 \Big[  \sum_{k=0}^{n-1}
          A_\mu^{(k)}\, d\hat \psi_k  \Big]dx_{\mu}\Big\}\,,\nn\\
\bar X_\mu &=& \sum_{k=0}^n c_k\, x_\mu^{2k}\,. \eea
It is straightforward to verify that $d\bar s^2$ is the metric for pure
(A)dS spacetime.  As in the odd dimensions, this metric can be put into
a diagonal form, while keeping the second term of (\ref{evenks1}) simple.
To do that, we first reparameterise $X_\mu$ as
\be
\bar X_\mu=-(1-g^2 x_\mu^2) \prod_{k=1}^{n-1}(a_k^2-x_\mu^2)\,.
\ee
We then complete the square in $d\bar s^2$, {\it i.e.}
\be  d\bar s^2 =   \sum_{\mu=1}^n \Big\{
            \fft{U_\mu}{\bar X_\mu}dx_{\mu}^2
          - \fft{\bar X_\mu}{U_\mu}
            \Big[  \sum_{k=0}^{n-1}
           A_\mu^{(k)}\, d\hat \psi_k
           - \fft{U_\mu}{\bar X_\mu} dx_{\mu}\Big]^2\Big\}
\ee
and make the coordinate transformation
\be d\td\psi_k = - d\hat \psi_k + \sum^n_{\mu=1}
                 \fft{(- x_\mu^2)^{n-k-1}}{\bar X_{\mu}}dx_{\mu}\,,
                 \qquad k \, = \, 0 \,,\:\cdots\:,n-1\,.
\ee
The metric (\ref{evenks1}) can then be put into a new form:
\be ds^2 = d\bar s^2 -
\sum_{\mu=1}^n\fft{2b_{\mu}x_{\mu}}{U_{\mu}}
           \Big[\sum_{k=0}^{n-1} A_\mu^{(k)}\, d\td \psi_k
           - \fft{U_\mu}{\bar X_\mu} dx_{\mu}\Big]^2\,,
\ee
where
\be  d\bar s^2 =   \sum_{\mu=1}^n \Big\{
            \fft{U_\mu}{\bar X_\mu}dx_{\mu}^2
          - \fft{\bar X_\mu}{U_\mu}
            \Big[  \sum_{k=0}^{n-1}
           A_\mu^{(k)}\, d\td \psi_k \Big]^2\Big\}\,.
\ee
The $d\bar s^2$ metric can now straightforwardly
be diagonalised by means of the coordinate transformation
\be     \tau =  \sum_{k=0}^{n-1} B^{(k)} \, d\td \psi_k\,,\qquad
  \fft{\varphi_i}{a_i} = \sum^{n-1}_{k=1}B^{(k-1)}_i d\td\psi_k
                 + g^2 \sum_{k=0}^{n-2} B_i^{(k)}\, d\td \psi_k
                \qquad i \, = \, 1 \,,\:\cdots\:,n-1\,,
           \ee
where
\be B_i^{(k)} = \sum'_{j_1 <j_2 <\cdots < j_k}
            a_{j_1}^2 a_{j_2}^2\cdots
        a_{j_k}^2\,,\qquad
        B^{(k)} =
         \sum_{j_1<j_2\cdots <j_k}
               a_{j_1}^2 a_{j_2}^2 \cdots a_{j_k}^2\,.
\ee
The even dimensional (A)-dS Kerr-NUT metrics can now be expressed as
\be ds^2 = d\bar{s}^2 - \sum_{\mu=1}^{n}
          \fft{2b_{\mu} x_{\mu}}{U_{\mu}}(k_{(\mu)\alpha}
dx^{\alpha})^2\,, \ee
where
\bea d\bar{s}^2 &=&  \fft{W}{\prod^{n-1}_{i=1} \Xi_i}
                      \,d\tau^2
            + \sum_{\mu=1}^{n} \fft{U_\mu}{\bar X_\mu }\, dx_\mu^2
            - \sum_{i=1}^{n-1} \fft{\gamma_i}
            {a_i^2 \Xi_i {{\prod}'}_{k=1}^{\, n-1} (a_i^2-a_k^2)}
            d\varphi_i^2\,, \\
 k_{(\mu)\alpha}dx^{\alpha} &=& \fft{W}{1- g^2 x_\mu^2}
                \fft{d\tau}{\prod^{n-1}_{i=1} \Xi_i}
                - \fft{U_{\mu} dx_{\mu}} {\bar X_\mu}
                - \sum_{i=1}^{n-1} 
\fft{\gamma_i d\varphi_i}
{(a_i^2- x_\mu^2) a_i \Xi_i 
{{\prod}'}_{k=1}^{\, n-1} (a_i^2-a_k^2)}\,,
\eea
where $\Xi_i$, $\gamma_i$ and $W$ have the same structure as that
in the even dimensions, given by (\ref{extrastr}).
When all but one of the $b_\mu$ vanishes, the metric reduces to the
Kerr-Schild form of the rotating (A)dS black hole obtained in
\cite{glpp}.

        To summarise, we find that in both even and odd dimensions,
the (A)dS-Kerr-NUT solution can be cast into the following
multi-Kerr-Schild form:
\be
ds^2=d\bar s^2 + \sum_{\mu=1}^n \fft{2b_\mu\, f(x_\mu)}{U_\mu}
(k_{(\mu)\alpha} dx^\alpha)^2\,,
\ee
where $f(x_\mu)=1$ for odd dimensions and $f(x_\mu)=x_\mu$ for
even dimensions.  The vectors $k_{(\mu)\alpha}$ are $n$
linearly-independent, mutually-orthogonal and affinely-parameterised
null geodesic congruences, satisfying
\be
k_{(\mu)\alpha} k^\alpha_{(\nu)} = 0\,, \qquad
k^\alpha_{(\mu)} \bar \nabla_\alpha k_{(\mu)\beta}=0
\,.
\ee
Note that the index $\alpha$ in $k_{\alpha(\mu)}$ can be raised
with either $g^{\alpha\beta}$ or $\bar g^{\alpha\beta}$ for
the above conditions to be satisfied.

\section{Harmonic 2-forms in $D=2n$ dimensions}

In this section, we find $n$ harmonic 2-forms $G_\2^{(\mu)}=
dB_\1^{(\mu)}$ on
the $2n$-dimensional (A)dS-Kerr-NUT metric (\ref{2npleb}),
where we use the index $\mu=1,2,\ldots n$ to label the harmonic
2-forms. The potentials have a rather simple form, given by
\be
B_\1^{(\mu)}=\fft{x_\mu}{U_\mu} \Big(\sum_{k=0}^{n-1}
A_\mu^{(k)}\, d\psi_k\Big)\,.
\ee
The metric (\ref{2npleb}) admits a natural vielbein basis, namely
\be
e^\mu=\fft{dx_\mu}{\sqrt{Q_\mu}}\,,\qquad
\td e^\mu=\sqrt{Q_\mu} \Big(\sum_{k=0}^{n-1}
A_\mu^{(k)}\, d\psi_k\Big)\,.\label{vielbein}
\ee
In this vielbein basis, the harmonic 2-forms $G_\2^{(\mu)}$ are given by
\be
G_\2^{(\mu)} = \sum f^{(\mu)}_\nu\, e^\nu\wedge \td e^\nu\,,
\label{g2form}
\ee
where the coefficients are
\bea
f^{(\mu)}_\mu &=& \fft{1}{U_\mu^2}\Big[A^{(n-1)} +
\sum_{k=1}^{n-2} (-1)^k (2k+1) x_\mu^{2(k+1)} A_\mu^{(n-k-2)}\Big]
\,,\nn\\
f^{(\mu)}_\nu &=& -\fft{2x_\mu x_\nu}{U_\mu^2}\prod_{\rho\ne \mu,\nu}
(x_\rho^2 - x_\mu^2)\,,\qquad \hbox{with $\mu\ne\nu$}\,.
\eea
We verify with low-lying examples that all of the $G_\2^{(\mu)}$ are
harmonic, {\it i.e.}~ $dG_\2^{(\mu)}=0 = d*G_\2^{(\mu)}$.
It is worth
observing that these 2-forms are harmonic regardless of the detailed
structure of the functions $X_\mu$.

        It was shown in \cite{clpnut} that the BPS limit of the metric
(\ref{2npleb}) gives rise to the non-compact Calabi-Yau metric that
can provide a resolutions of the cone over the Einstein-Sasaki
spaces.  Under suitable coordinate transformation, the metric is given
by
\be ds^2= \sum_{\mu=1}^n \Big\{
         \fft{dx_\mu^2}{Q_\mu} + Q_\mu\, \Big( \sum_{k=0}^{n-1}
          A_\mu^{(k)}\, d\psi_k\Big)^2\Big\}\,,\label{2nplebbps}
\ee
where we define
\bea Q_\mu &=& \fft{4X_\mu}{U_\mu}\,,\qquad U_\mu =
{{\prod}'}_{\nu=1}^n
     (x_\nu - x_\mu)\,, \qquad X_\mu = x_\mu
\prod_{k=1}^{n-1} (x_\mu + \alpha_k) + 2 b_\mu\,,\nn\\
A_\mu^{(k)} &=& \sum'_{\nu_1 <\nu_2 <\cdots < \nu_k}
  x_{\nu_1} x_{\nu_2}\cdots
        x_{\nu_k}\,.\label{QYevenbps}
\eea
Note that we have Wick rotated the metric to have Euclidean signature.
We can choose the same form of the vielbein basis as in (\ref{vielbein}).
The K\"ahler 2-form is then given by
\be
J_\2=\sum_{\mu=1}^n
e^{\mu}\wedge\td e^{\mu}\,.
\ee
The 1-form potentials for the harmonic 2-forms are given by
\be
B_\1^{(\mu)} = \fft{1}{U_\mu}\Big(\sum_{k=0}^{n-1} A_\mu^{(k)}\,
d\psi_k \Big)\,.
\ee
The corresponding harmonic 2-forms $G_\2^{(\mu)}$ have the same form
as in (\ref{g2form}), with the functions $f_\nu^{(\mu)}$ are given by
\be
f_\nu^{(\mu)} = \fft{2}{U_\mu^2}
\prod_{\rho\ne \mu,\nu} (x_\rho - x_\mu)\,,
\hbox{with $\mu\ne \nu$}\,,\qquad
f_\mu^{(\mu)} = - \sum_{\nu\ne\mu} f_\nu^{(\mu)}\,.
\ee
Note that $G_\2^{(\mu)}$ satisfy the linear relation
$\sum_{\mu=1}^n G_\2^{(\mu)}=0$.
Thus, in the BPS limit, there are $(n-1)$ linearly independent
such harmonic 2-forms.  Together with the K\"ahler 2-form,
the total number of harmonic 2-forms is $n$ again.

\section{Conclusion}

         In this letter, we explicitly express the general
(A)dS-Kerr-NUT metrics in Kerr-Schild form for both even and odd
dimensions.  We demonstrate that, in a suitable coordinate system the
mass, NUT and angular momentum parameters enter linearly in the metric,
and hence they can be viewed as a linear perturbation of pure
(A)dS spacetime.

         We also obtain $n$ harmonic 2-forms on the $2n$-dimensional
(A)dS-Kerr-NUT metrics.  An interesting property of these harmonic
2-forms is that the closure and co-closure do not depend on the
detailed structure of the functions $X_\mu$. This provides a potential
ansatz for charged (A)dS-Kerr-NUT solutions for pure Einstein-Maxwell
theories in higher dimensions, whose explicit analytical solutions
remain elusive.  In the case of four dimensions, the back-reaction of
the gauge field to the Einstein equations gives precisely the charged
Plebanski metric \cite{pleb}, where only the functions $X_\mu$ in the
metric have extra contributions from the electric and magnetic
charges.  However, the same phenomenon does not occur in higher
dimensions; nevertheless, the harmonic 2-forms we constructed can be
viewed as charged (A)dS-Kerr-NUT solutions at the linear level for
small-charge expansion.  Together with the charged slowly-rotating
black holes obtained in \cite{aliev1,aliev2}, our results may
lead to the general charged (A)dS-Kerr-NUT solutions.

\section*{Acknowledgement}

      We are grateful to Chris Pope and Justin V\'azquez-Poritz
for useful discussions.


\begin{thebibliography}{99}

\bm{ksform} R.P. Kerr and A. Schild, {\it Some algebraically degenerate
solutions of Einstein's gravitational field equations,}
Proc. Symp. Appl. Math. {\bf 17}, 199 (1965).

\bm{pleb} J.F. Plebanski, {\it A class of solutions of Einstein-Maxwell
equations}, Ann. Phys. {\bf 90}, 196 (1975).

\bm{cglp} Z.W. Chong, G.W. Gibbons, H. L\"u and C.N. Pope,
{\it Separability and killing tensors in Kerr-Taub-NUT-de sitter metrics in
higher dimensions,} Phys.\ Lett.\ {\bf B609}, 124 (2005), hep-th/0405061.

\bm{clpnut} W. Chen, H. L\"u and C.N. Pope,
{\it General Kerr-NUT-AdS metrics in all dimensions},
  Class.\ Quant.\ Grav.\  {\bf 23}, 5323 (2006), hep-th/0604125.

\bm{curvature} N. Hamamoto, T. Houri, T. Oota and Y. Yasui,
{\it Kerr-NUT-de Sitter curvature in all dimensions,}
  J.\ Phys.\  {\bf A40}, F177 (2007), hep-th/0611285.

\bibitem{classification} A. Coley, R. Milson, V. Pravda and A. Pravdova,
{\it Classification of the Weyl tensor in higher-dimensions,}
  Class.\ Quant.\ Grav.\  {\bf 21}, L35 (2004), gr-qc/0401008.

\bibitem{separability} V.P. Frolov, P. Krtous and D. Kubiznak,
{\it Separability of Hamilton-Jacobi and Klein-Gordon equations in general
  Kerr-NUT-AdS spacetimes,}
  JHEP {\bf 0702}, 005 (2007), hep-th/0611245.

\bibitem{killingyano} P. Krtous, D. Kubiznak, D.N. Page and V.P. Frolov,
{\it Killing-Yano tensors, rank-2 Killing tensors, and
conserved quantities in
  higher dimensions,} JHEP {\bf 0702}, 004 (2007), hep-th/0612029.

\bibitem{ypq} J.P. Gauntlett, D. Martelli, J. Sparks and D. Waldram,
{\it Sasaki-Einstein metrics on $S^2 \times S^3$,}
  Adv.\ Theor.\ Math.\ Phys.\  {\bf 8}, 711 (2004), hep-th/0403002.

\bm{lpqr} M. Cveti\v c, H. L\"u, D.N. Page and C.N. Pope,
{\it New Einstein-Sasaki spaces in five and higher dimensions,}
  Phys.\ Rev.\ Lett.\  {\bf 95}, 071101 (2005), hep-th/0504225.

\bibitem{res1} T. Oota and Y. Yasui,
{\it Explicit toric metric on resolved Calabi-Yau cone,}
  Phys.\ Lett.\  {\bf B639}, 54 (2006), hep-th/0605129.

\bibitem{res2} H. L\"u and C.N. Pope,
{\it Resolutions of cones over Einstein-Sasaki spaces,},
hep-th/0605222.

\bibitem{glpp} G.W. Gibbons, H. L\"u, D.N. Page and C.N. Pope,
{\it The general Kerr-de Sitter metrics in all dimensions,}
  J.\ Geom.\ Phys.\  {\bf 53}, 49 (2005), hep-th/0404008.

\bibitem{aliev1} A.N. Aliev,
{\it Charged Slowly Rotating Black Holes in Five Dimensions,}
  Mod.\ Phys.\ Lett.\  {\bf A21}, 751 (2006), gr-qc/0505003.

\bibitem{aliev2} A.N. Aliev,
{\it Rotating black holes in higher dimensional Einstein-Maxwell gravity,}
  Phys.\ Rev.\  {\bf D74}, 024011 (2006), hep-th/0604207.

\end{thebibliography}
\end{document}